\documentclass[aps,prd,twocolumn,floatfix]{revtex4}
\usepackage{amsmath,float}

\usepackage{graphicx}
\usepackage{dcolumn}
\usepackage{bm}
\usepackage{array}
\usepackage{color}

\begin{document}

\title{Formation and evaporation of strangelets\\during the merger of two compact stars}

\author{Niccol\'o Bucciantini}

\affiliation{INAF - Osservatorio Astrofisico di Arcetri, Largo E. Fermi 5, 50125 Firenze, Italy}
\affiliation{Dipartimento di Fisica e Astronomia, Universit\`a degli Studi di Firenze, Via G. Sansone 1, 50019 Sesto F. no (Firenze), Italy}
\affiliation{INFN - Sezione di Firenze, Via G. Sansone 1, 50019 Sesto F. no (Firenze), Italy}

\author{Alessandro Drago}
\author{Giuseppe Pagliara}
\author{Silvia Traversi}
\affiliation{Dipartimento di Fisica e Scienze della Terra, Universit\'a di Ferrara, Via Saragat, 1, 44122 Ferrara, Italy}
\affiliation{INFN - Sezione di Ferrara, Via Saragat, 1, 44122 Ferrara, Italy}

\author{Andreas Bauswein}
\affiliation{GSI  Helmholtzzentrum  f\"ur Schwerionenforschung, Planckstrassee 1, 64291  Darmstadt, Germany}  
\affiliation{Helmholtz Research Academy Hesse for FAIR (HFHF), GSI Helmholtz Center for Heavy Ion Research, Campus Darmstadt, Germany}

\begin{abstract}
We study the partial fragmentation of a strange quark star into strangelets during the process of merger of two strange quark stars. We discuss the fate of the fragments considering their possible evaporation into nucleons. We show that only a rather small amount of large size strangelets, ejected from the spiral arms in the post-merger, survives a total evaporation into nucleons. 
In this way we demonstrate that: 1) the density of strangelets in the galaxy is too low to trigger the conversion of all neutron stars into strange quark stars and it allows the co-existence of both types of compact objects; 2) the probability of direct detection of a strangelet is negligible and therefore its non-detection is compatible with the strange quark matter hypothesis; 3) most of the matter ejected during and after the merger of two strange quark stars evaporates into nucleons and therefore it can generate a kilonova-like signal. 
\end{abstract}

\maketitle 

\section{Introduction}
\label{sec:level1}
One of the oldest and most relevant criticisms of the existence of Strange Quark Stars \cite{Witten:1984rs,Haensel:1986qb} (QSs), i.e. self-bound objects made of absolutely stable strange quark matter, 
is that if QSs exist then neutron stars (NSs) would not exist. The density of strangelets, i.e. of fragments of strange quark matter produced e.g. in a merger of two QSs, would be large enough that at least one fragment would be captured by every NS (or by their progenitors) forcing the entire star to deconfine and to transform into a QS \cite{Madsen:1989pg}. Another quite obvious objection concerns the absence of strangelets detected in experiments: if Witten's hypothesis \cite{Witten:1984rs} is correct, one would expect not only that QSs can exist, but also that strangelets can be stable and with a mass significantly smaller than that of a star as a result of the self-binding of strange quark matter. Finally, the kilonova signal AT2017gfo \cite{Arcavi:2017xiz,Tanvir:2013pia,Kasen:2017sxr} may be in tension with the existence of QSs. The signal was generated by the formation and decay of heavy nuclei produced from a gas of nucleons ejected from the external layer of the merging stars. This either implies necessarily the existence of NSs or it requires a mechanism to reconvert strange matter to nuclear matter, which could then produce the observed signal. Instead, if QSs cannot co-exist with NSs and if nucleons cannot be abundantly produced in mergers involving QSs, then Witten's hypothesis has to be excluded.

While strangelets can rather easily be produced by QS mergers, their fate is unclear, since they can evaporate if the temperature is sufficiently high. The possibility of strangelet evaporation has been studied in a cosmological context in \cite{Alcock:1985vc,Madsen:1986jg}, 
where it was shown that only lumps with a large baryon number can survive.
Here, we will adopt the formalism developed in \cite{Alcock:1985vc} and we will take into account the criticisms of \cite{Madsen:1986jg} 
but we will consider a different scenario, i.e. the production of strangelets at the moment of the merger of two compact stars (cf.~\cite{Paulucci:2014vna,Biswas2017,Lai2020}). Traditionally, this problem has been associated with the merger of two QSs \cite{Bauswein:2008gx,Bauswein:2009im,Paulucci:2014vna} \footnote{The merger of a QS with a black-hole was discussed in \cite{Kluzniak:2002dm} where no mass ejection was observed.}, but in the last years it has been suggested that QSs and NSs can coexist \cite{Drago:2013fsa,Drago:2015cea,Drago:2015dea,Wiktorowicz:2017swq}. 
In this regard we emphasize that tidal deformability constraints as well as the potential merger outcome are compatible with GW170817 being a QS-QS or QS-NS merger~\cite{Drago:2017bnf,Burgio:2018yix,DePietri:2019khb}.

In this paper, by using the theory of fragmentation and results of merger simulations, we show that only very massive fragments of quark matter can survive evaporation. Therefore, their density in the galaxy is low enough to allow the co-existence of NSs with QSs. This same argument also clarifies why strangelets have never been detected in experiments. Finally, we show that most of the quark matter constituting strangelets evaporates into nucleons and that this process takes place close to the central region of the merger. Therefore, the evolution of the evaporated material should be similar to that of the nucleonic material ejected during the merger of two NSs and thus it can generate a kilonova-like signal.

\section{Ejection of quark matter}
\label{sec:level4}
QSs can in principle eject significant amounts of matter during the merger with another QS or with a NS. In NS-NS mergers matter is ejected dynamically from the collision interface and by tidal disruption, and on secular time scales from the disk surrounding the central remnant, e.g.~\cite{Hotokezaka:2012ze,Bauswein:2013yna,Radice:2016dwd,Fernandez2016}.  There exists only a limited number of hydrodynamical simulations of QS-QS mergers \cite{Bauswein:2008gx,Bauswein:2009im,Zhu:2021xlu,Zhou:2021tgo}.
Here we use the results of the simulations of Refs. \cite{Bauswein:2008gx,Bauswein:2009im}, concentrating on the trajectories of fluid elements which are ejected from the merger site.
In particular, we refer
to the simulation having a bag constant $B=60 \mathrm{MeV/fm}^3$ for which the equilibrium density (at which the pressure vanishes) is $\rho_{eq}\sim 4.5\times 10^{14} \mathrm{g/cm}^3$.
In Fig.\ref{trajectory} we display an example of trajectory of a particle ejected: its density, its temperature and the change in its scalar velocity are shown as a function of time. 

No simulations were available for NS-QS mergers, and here we briefly discuss the first calculations for these systems. A short summary of the simulations of a NS-QS merger is provided in Appendix A. The main results are: 
\begin{itemize}
\item 
In QS-QS mergers matter coming mainly from the collision interface 
moves towards the exterior region of the remnant; from this area thin spiral arms form and strange matter is ejected from the tips of those rapidly rotating arms (Fig.~4 of \cite{Bauswein:2009im}).
\item For NS-QS mergers, almost all the ejected matter originates from the NS with no significant amount of unbound quark matter until the end of the simulations. 
This is because the QS is strongly bound and disrupts the NS during the collision. The nuclear material is wrapped around the QS preventing the ejection of strongly bound quark matter.  In any case, even if some unbound quark matter was produced, the thermodynamical conditions of the ejected quark matter would be very similar to those in QS-QS mergers.
\end{itemize}

No simulations of the secular evolution of QS-QS or QS-NS merger remnants exist, but we expect that no quark matter is ejected from those systems beyond dynamical time scales. Macroscopic chunks of self-bound quark matter orbiting the central remnant (Fig.~4 of \cite{Bauswein:2009im}) behave approximately like test-particles in contrast to the hydrodynamical evolution and viscous secular processes in tori of NS-NS mergers. Hence it appears very unlikely that secular processes could unbind fragments of quark matter. Evaporation from gravitationally bound quark matter fragments may be possible but does not contribute to the strangelet production.

\section{Fragmentation of quark matter}
\label{sec:level5}
 Here and in the the following we will concentrate on the QS-QS case. In that type of mergers, as shown in Ref.\cite{Bauswein:2009im}, matter is ejected by the fragmentation of the spiral arms, like water drops ejected from a rotating sprinkler or droplets foaming from ocean waves.
The final size of the fragments (before evaporation) depends on the initial fragmentation of bulk quark matter and the effect of rescattering. The initial size of these fragments is regulated by turbulence 
\citep{kolmogorov,hinze,frohn,lefebvre1988atomization}: forces acting on large scales (e.g. tidal forces),  as well as  large scale perturbations like spiral arms, can drive turbulence, which cascades to smaller scales causing turbulent fragmentation. 

\begin{figure}[t]
\includegraphics[width=10.5cm]{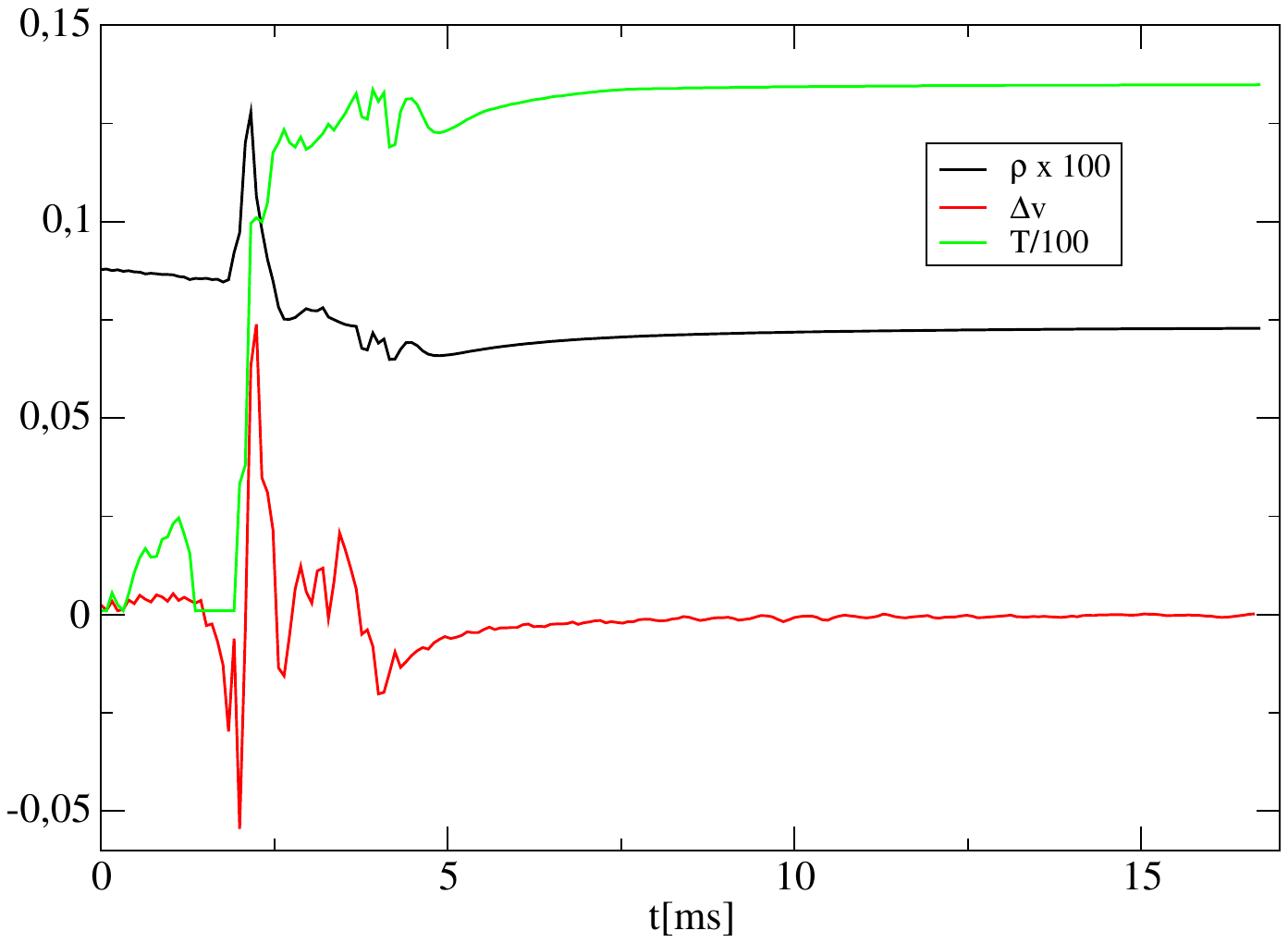}
\caption{Example of the trajectory of an ejected particle. Density (in $M_\odot ^{-2})$, temperature (in MeV) and changes in the scalar velocity (in units of c) tested on a time-scale of 0.0788~ms.}\label{trajectory}
\end{figure}

 The Kolmogorov scale is defined in terms of quark matter shear viscosity $\mu$ \cite{Heiselberg:1993cr}, density $\rho$, and energy dissipation rate $\epsilon$, as $l_{\rm K} = (\mu^3/\rho^3 \epsilon)^{1/4}$. Above $l_{\rm K}$ the turbulent  energy dissipation rate is constant (inertial range), while at $l_{\rm K}$ the turbulent cascade enters the visco-dissipative regime. Using smoothed particles hydrodynamics simulations \cite{Bauswein:2008gx,Bauswein:2009im}, we estimate a turbulent velocity on the scale $R$ (of the order of the size of the particles of the simulations), $v(R)\sim (0.01-0.03)c$,  see the red-line of Fig.\ref{trajectory}. The energy dissipation rate in the inertial range is estimated as $\epsilon\sim v(R)^3/R$ \cite{tennekes,davidson}.
Fragmentation proceeds down to smaller scales  until either one reaches  the dissipation scale or the surface tension limit. The relative importance of surface tension is parametrized by the size-dependent Weber number, defined as the ratio between turbulent kinetic energy and surface tension, $\mathrm{We}(d)=(\rho/\sigma) v(d)^2 d$, where  $\sigma$ is the surface tension and $v(d)$ the turbulent velocity at the scale $d$. Surface tension halts turbulent fragmentation when $\mathrm{We}(d)\simeq O(1)$. In particular, experimental results suggest that fragmentation happens only if $\mathrm{We}(d)\gtrsim 4$ (see e.g. \cite{deane2002scale} and references therein), and this allows for the definition of a minimum size of droplets, called Hinze scale $l_H$.

For matter ejected from the spiral arms in the QS-QS merger we set $\sigma\sim 10 \,\mathrm{MeV/fm^2}$ and we adopt for the viscosity 
the equation:
\begin{equation}
\frac{\mu}{\mathrm{g\,cm^{-1}\,s^{-1}}}  =1.7\times 10^{18} \left(\frac{0.1}{\alpha_s}\right)^{5/3} \rho_{15}^{14/9}T_9^{-5/3} 
\end{equation}
where $\rho_{15}=\rho/10^{15} \mathrm{g\,cm^{-3}}$
and $T_9=T/(10^9K)$ \cite{Heiselberg1993}.
With $\alpha_s=0.1$, $\rho=5\times 10^{14}$ g cm$^{-3}$ and $T=10^{11}$K
one finds that $l_K\sim 3\times 10^{-6}$cm and
$l_H \sim 5\times 10^{-5}$cm, showing that turbulent fragmentation is actually limited by surface tension and that the smallest droplets corresponds to a minimum baryon number $A_H\approx 10^{26}$. 

These droplets will rescatter many times immediately after the ejection (their mean free path in the region close to the merger is  $\sim 10^{-3}$ cm assuming a spherical shell of radius $10$~km and width $1$~km) and they will  undergo further fragmentation, now due to collisional interactions. Collisional fragmentation will proceed by producing smaller sized fragments, until either the surface tension limit is reached (which can again be defined through the Weber number $\mathrm{We}(d)\gtrsim O(1)$, taking now the scale $d$ as the droplet size and the velocity $v(d)$ as the typical impact speed), or the viscosity limit is reached, at which viscous dissipation suppresses fragmentation. The relative importance of viscosity with respect to surface tension is given by  the size dependent Ohnesorge number, $\mathrm{Oh}(d)=\mu/(\rho\sigma d)^{1/2}$:  $\mathrm{Oh}(d) \simeq 1$ sets the boundary between viscous limited and tension limited fragmetation (see e.g. Fig. 1 of \cite{hsiang} and \cite{lefebvre1988atomization}). 
As we will show, most of the ejected material passes through a violent shock and reaches temperatures larger than about 15 MeV. Under those conditions viscosity plays a marginal role and fragmentation is actually limited again by surface tension.

\begin{figure}[t]
\includegraphics[width=10.5cm]{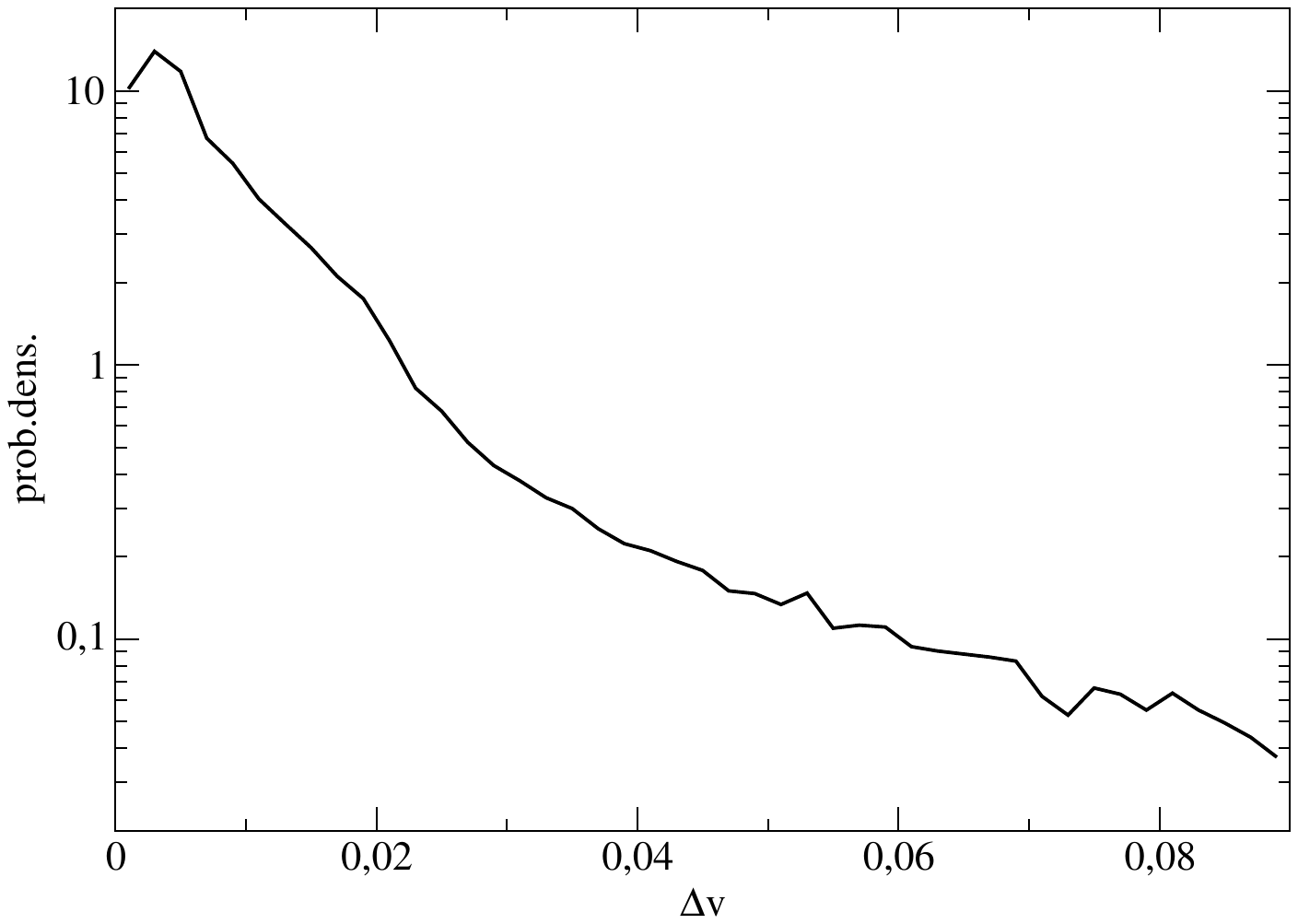}
\caption{Distribution of the changes in the scalar velocity of the particles (about 1 km size) in units of $c$}\label{velocities}
\end{figure}

In order to understand the fate of the material ejected during and after the merger we have analyzed the trajectories of the particles using the results of the simulations described in \cite{Bauswein:2008gx,Bauswein:2009im}. All the ejected particles display a violent shock at the moment of the first contact between the two QSs, signalled by the large and rapid change of the scalar velocity,  see Fig.\ref{velocities} where we show the distribution of the changes in the velocity \footnote{Clearly these changes depend on the time-scale on which they are tested (in the analysis presented here the velocities are measured every 0.0788 ms), but the purpose of our simulation is only to show an example of fragmentation due to rescattering without any ambition of precision. Notice that the scale of the collisional velocities estimated in this way turns out to be similar
to the turbulent velocities in a neutron star - neutron star merger \cite{Giacomazzo:2014qba,Kiuchi:2015sga}.
}. At the same time, the temperature increases 
whereas, in most cases, the density decreases immediately after the shock, indicating that the material is decompressed (see the example displayed in Fig.\ref{trajectory}). 
This suggests that violent collisions among the particles occur, which generate fragmentation into small pieces. During this period densities below the equilibrium one (at which the pressure vanishes) are reached and this is the optimal condition for the system to fragment.

\begin{figure}[b]
\includegraphics[width=\linewidth]{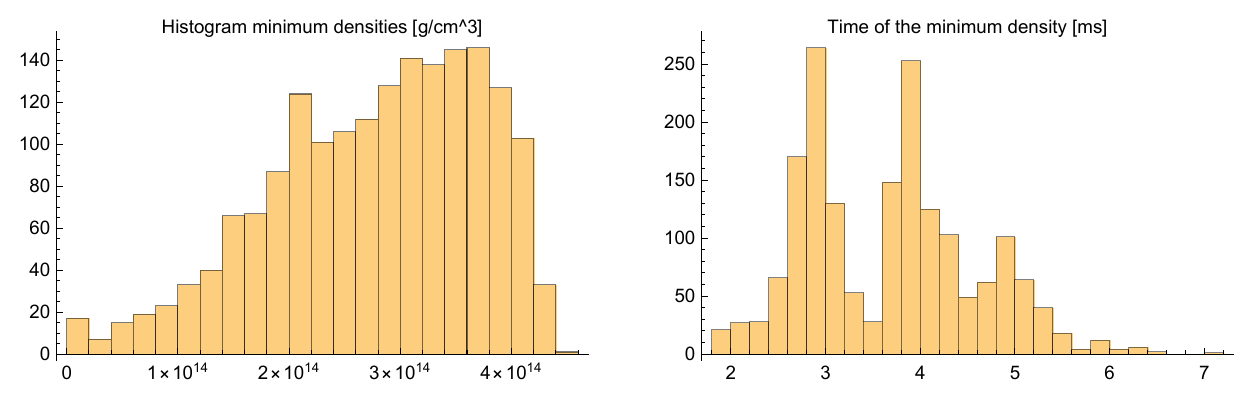}
\caption{Left panel: distribution of the minimum densities reached by the particles during their evolution. Right panel: distribution of the times at which the minimum densities of the left panel are reached.}\label{histos}
\end{figure}

 Clearly, in a hydrodynamical simulation it is impossible to explore the regime of the spinodal instability which takes place at much lower densities and at which spontaneous fragmentation would start, but we can investigate which are the minimum densities reached during the evolution of each of the particles and when those densities have been reached. We display the results in Figs.~\ref{histos} and we can notice that the minimum densities of most of the particles are significantly lower than $\rho_{eq}$ and that they are reached soon after the first collision (which takes place at about 2.5 ms). Interestingly, we can notice the hint of a periodicity, with a period of about 1 ms, suggesting that during the first milliseconds of the evolution the system oscillates and that during those oscillations low densities are reached. Also, these first milliseconds after the first touch of the QSs are exactly the 
moment during which we have noticed the violent changes in the velocities displayed in Fig.~\ref{velocities} and we therefore assume that it is during this period that fragmentation takes place.

 \subsection{Simulation of scattering}
\begin{figure}[t]
\includegraphics[width=\linewidth]{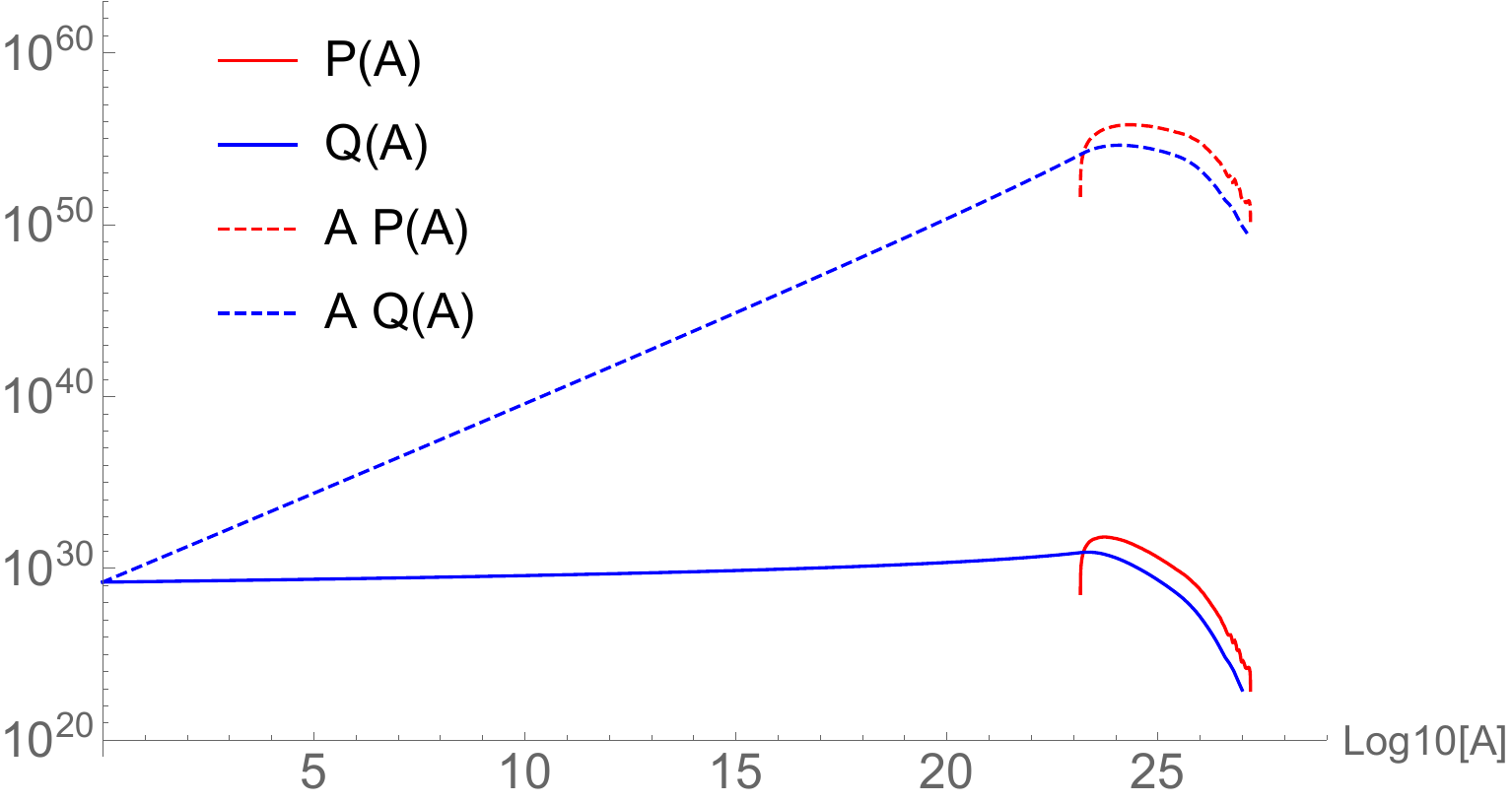}
\caption{Number density of fragments before, $P(A)$, and after evaporation, $Q(A)$, as functions of $\mathrm{Log}_{10} A$. Also shown are the corresponding distributions of the baryonic mass of the fragments. The normalization is chosen to be $A_\mathrm{tot}=\int_1^\infty \mathrm{d}A\, A P(A)=10^{56}$ (see text).}\label{distributions}
\end{figure}

We have performed a MonteCarlo simulation sampling the distribution of the collisional velocities discussed above and displayed in Fig.~\ref{velocities}. The initial size of the fragments was the same as the size of the particles, i.e. about 1 km
and it corresponds to $A_\mathrm{init}\sim 10^{43}$. The estimate of the distribution of the final size of the fragments was based on the following algorithm:
\begin{itemize}
\item generation of a new collisional velocity;
\item scaling of the velocity down to the size of the fragment undergoing the collision by using Kolmogorov scaling;
\item evaluation of $\mathrm{We}$ at the new scale;
\item breaking of the fragment into a number of pieces that is assumed to be equal to $\mathrm{We}$ as long as $\mathrm{We} > 4$. This procedure is a very simplified implementation of the idea that the number of fragments increases with $\mathrm{We}$ \cite{Ashgriz:1990}. In reality, the number of fragments forming in each collision depends also on the impact parameter and ours is possibly an overestimate of the number of fragments, but in this way we can at least put some lower limit to the minimum size of the fragments formed after the re-scattering and before evaporation.
\end{itemize}
This procedure is iterated a few ten times, similar to the number of large changes in the scalar velocity of each particle observed in the simulation of the merger, and it effectively halts because the last iterations produce fragments so small that $\mathrm{We}\lesssim 4$. The outcome of this chain of scatterings are $A_\mathrm{init}/A_\mathrm{final}$ fragments all having baryon number $A_\mathrm{final}$.  We have simulated $10^6$ chains of collisions and we have obtained $P(A)$, the distribution of the size of the fragments displayed in Fig. \ref{distributions}. Notice that the distribution is centered at about $A\sim 10^{24}-10^{25}$, a number only slightly smaller than the one estimated by using the turbulent velocities of a neutron star - neutron star merger. 
The sharp lower limit of $P(A)$ is regulated by the maximum energy available for fragmenting strangelets: a much larger energy would be needed to obtain significantly smaller fragments. Notice that smaller fragments produced by the re-scattering still have a size larger or comparable with the $l_K$, justifying the usage of Kolmogorov scaling in our simulation. 
With this procedure not only we have estimated the average size of the fragments, but we have also obtained a distribution law of their sizes, and we will use that distribution in the evaluation of the evaporation of the fragments.
In order to check the stability of our results with respect to the assumptions made on the collisional velocities we have repeated the simulation by arbitrarily assuming that the velocities are larger by a factor of 2. In this way we obtain a number of fragments that is roughly one order of magnitude larger, but still very far from causing problems from the phenomenological viewpoint. It is very important to notice that the evaporation of the fragments, partial or total, can only reduce their number, which is therefore limited by the number of fragments produced before the evaporation which, in turns, depends on the energy available in the collisions. This simple but crucial argument was not taken into account in previous analyses.

\section{Evaporation of strangelets} 
Following \cite{Alcock:1985vc}, the evaporation consists in the emission of a baryon (usually a neutron) leaving a strangelet with baryon number $A-1$. This kind of process is endothermic and it requires an amount of energy equal at least to the binding (or ionization) energy $I$. In the following we concentrate on the range $I\sim (50-70) \mathrm{MeV}$, since these values correspond to the ones needed to stabilize QSs up to about 2 $M_\odot$ \cite{Drago:2019tbs}.  A first estimate of the evaporation rate stems from the detailed balance principle applied to the equilibrium situation in which evaporation and the re-absorption rates are equal. The evaporation rate into nucleons reads \cite{Alcock:1985vc}: 
 \begin{equation}
     \frac{dA}{dt}=\frac{1}{2\pi^2}m_nT^2e^{-\frac{I}{T}}(f_n+f_p)\sigma_0A^{\frac{2}{3}}
     \label{rate}
 \end{equation}
 where $m_n$ is the nucleon mass, $T$ is the temperature and $f_{n,p}$ are absorption efficiencies (for neutrons and protons) correcting the geometric cross section $\sigma_0 A^{\frac{2}{3}}$. 
 
 The simple evaporation rate of Eq.~(\ref{rate}) needs to be modified by taking into account two physical processes: the cooling of the strangelet due to evaporation and the possible absorption of neutrons from the environment  (i.e. one needs to compute the difference between evaporation and absorption rates, characterizing and off-equilibrium situation). The scenario we are considering is different from the one discussed in \cite{Alcock:1985vc,Madsen:1986jg} where the fate of strangelets produced during the cosmological baryogenesis was analyzed. In that situation the evolution of the temperature depends on the expansion rate of the Universe while in our case it is determined by the expansion and the cooling of the ejected material as estimated in the simulation of the merger. In the case of QS-QS mergers the local density of nucleons is determined by the evaporation process itself, as in the cosmological case. Instead, when discussing evaporation in a NS-QS merger one should also consider the density of nucleons already present in the system.

Because of the evaporation, the temperature of the strangelet $T_s$ is always smaller than the environment temperature $T_u$, but in order to have a significant evaporation rate, $T_s$ needs to remain comparable to $I$. The most basic mechanism to re-heat the strangelet is based on neutrinos:
$T_s$ is determined by an equilibrium condition between the energy lost by the strangelet, both because of the evaporation and of the neutrino emission, and the energy provided to the strangelet by the neutrino absorption (see Eqs.~(15,22) of \cite{Alcock:1985vc}):
\begin{equation}
    4\pi r_s^2 \bigl[\frac{7\pi^2}{160} \bigr][T_u^4 p(r_s,T_u)-T_s^4 p(r_s,T_s)]=\frac{dA}{dt}(I+2T_s)
    \label{neutrini}
\end{equation}
where $r_s$ is the radius of the strangelet and $p(r_s,T)$ is the probability of interaction between the neutrino and the strangelet. Here the net evaporation rate ${dA}/{dt}$ is the difference between the rates of evaporation and absorption:
\begin{equation}
    \frac{dA}{dt}=\bigl[\frac{m_nT_s^2}{2\pi^2}e^{-\frac{I}{T_s}} - N_n\bigl(\frac{T_s}{2\pi m_n}\bigr)^{\frac{1}{2}}\bigr](f_n+f_p)\sigma_0A^{\frac{2}{3}}
    \label{evap}
\end{equation}
where $N_n$ is the nucleon density.
By solving Eqs.~(\ref{neutrini},\ref{evap}) one can estimate the time-scale of evaporation of a strangelet of baryon number $A$ as a function of the density and temperature of the environment.

\begin{figure}
\includegraphics[width=\linewidth]{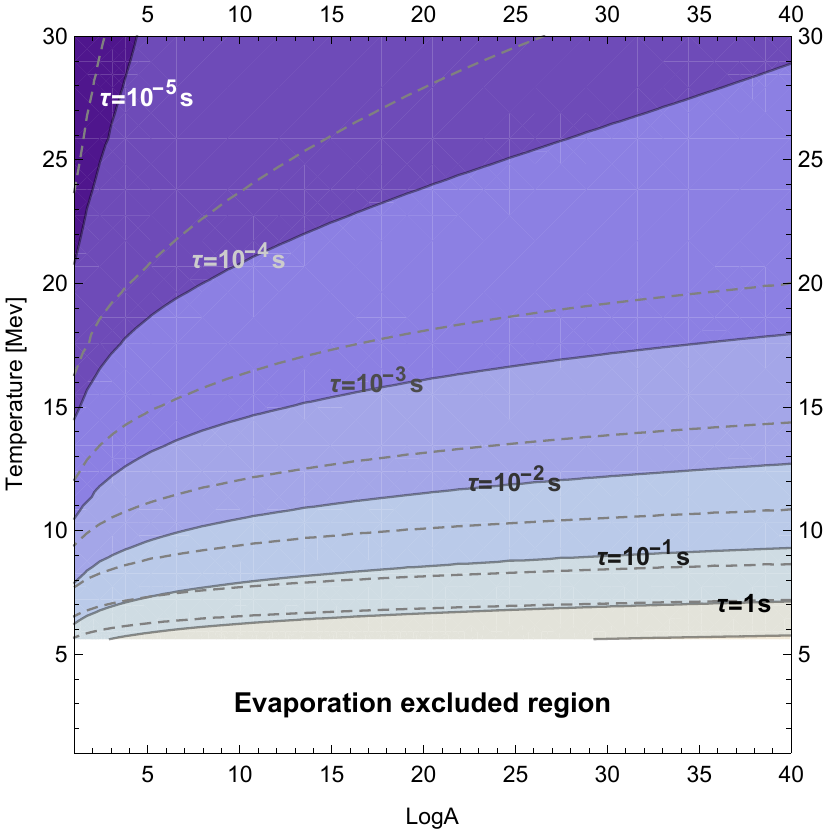}
\caption{Evaporation time-scale computed by assuming that neutrino absorption is the only re-heating mechanism and that the nucleon density is determined by the evaporated nucleons. Solid lines and color shading refer to $I=50\, \mathrm{MeV}$, the dashed lines correspond to $I=70\, \mathrm{MeV}$.}\label{timescaleevap}
\end{figure}

We solve the previous equations in the case relevant for QS-QS merger, i.e. a nucleon density generated only by the evaporation of strangelets. This density, to be included in Eq.~(\ref{evap}) in place of the environment density $N_n$, can be calculated by imposing the mechanical equilibrium between the components of the system, by considering the pressures due to electrons and photons and the ones due to the ejected neutrons and protons \cite{Alcock:1985vc}:
$N_n(T_s)=11\pi^2(T_u^4-T_s^4)/(360T_s)$.
Fig.~\ref{timescaleevap} displays the evaporation timescales as functions of the temperature and of the baryon number. The white area indicates that for $T\lesssim 5.6\, \mathrm{MeV}$ evaporation never takes place because the re-absorption rate always overcomes the evaporation rate. It is possible to demonstrate  (see Appendix B) that under the conditions discussed above the evaporation time-scale: 
\begin{equation}
\tau_A\propto \mathrm{Log} A.
\label{scaling}
\end{equation}

As discussed in \cite{Madsen:1986jg} the rapid evaporation of strangelets into nucleons 
enhances the strangeness fraction (specially close to the surface of the strangelet) and can make it energetically not favorable to further proceed with the evaporation unless weak reactions have the time to re-equilibrate the system. This imposes an upper limit on the evaporation rate. Following \cite{Alcock:1985vc,Madsen:1989pg} we incorporate this constraint, also taking into account Pauli blocking and temperature dependence: as estimated in \cite{Heiselberg:1986pg} the weak reaction rates are suppressed by 3--4 orders of magnitude with respect to the unquenched rate used in \cite{Alcock:1985vc} and the limit becomes:
$   dA/dt\lesssim K_{\mathrm{PB}}\, G_F^2\,\mu_q^5\,\sin^2\theta_c\,A$,
where $K_{\mathrm {PB}}\sim 10^{-3}-10^{-4}$ and $\mu_q$ is the chemical potential of the s quarks.
This limit on the evaporation time-scale depends again only logarithmically on $A$ and implies that the evaporation cannot be completed in less than about $(10^{-6}-10^{-5})\,$s even for light fragments. Since in our system the relevant time-scale is larger than at least $10^{-4}$s  (on a shorter time scale matter does not move significantly), the limit stemming from the weak reaction rate does not affect our results. 

To discuss the at least partial evaporation of the fragments we investigate the temperatures reached by the ejected particles along their trajectories: most of the particles reach temperatures exceeding 10 MeV and a significant fraction of quark matter 
evaporates into nucleons. We estimate the fraction of evaporated matter by using Eq.~(\ref{scaling}) and by comparing the time needed for that partial evaporation with the time each of the particles spends above a given temperature. In this way we obtain $Q(A)$, the distribution of fragment sizes after evaporation displayed in Fig.~\ref{distributions}.
See Appendix C for more details.

\section{Phenomenological implications and conclusions} 

We discuss now the phenomenological implications of the existence of strangelets produced in QS-QS mergers. 

If we normalize the total amount of matter ejected during a merger before evaporation as $\int_1^\infty {\mathrm d}A \,A\, P(A) = A_\mathrm{tot}$, then after evaporation we obtain $\int_1^\infty {\mathrm d}A \,A\, Q(A) \simeq 0.07 A_\mathrm{tot}$, indicating that about $93\%$ of the mass ejected as strangelets evaporated into nucleons. This evaporation takes place within at most 15 ms (the duration of the simulation) and therefore this very neutron rich material is similar to the one produced in NS-NS mergers and can synthesize heavy nuclei, which generate a kilonova. 

Assuming that the number of Galactic QS-QS mergers is $\sim 10$ per billion years~\cite{Wiktorowicz:2017swq} and that each event ejects $\sim 0.01 M_\odot$, the total number of ejected baryons is $~\sim A_\mathrm{tot}^\mathrm{gal} = 10^{56}$ and the total number of strangelets in the Galaxy is $N_\mathrm{tot}^\mathrm{gal}=\int_1^\infty {\mathrm d}A\, Q(A) = 3\times 10^{31}$, roughly half of $\int_1^\infty {\mathrm d}A \,P(A)$ because half of the strangelets evaporated completely. 
We can now answer the question posed in~\cite{Madsen:1989pg}  if the flux of strangelets is large enough so that one of them can be trapped into a star. The rate of impact of strangelets having a baryon number between $A_1$ and $A_2$ onto a star having mass M and radius R is \cite{Madsen:1989pg}:
\begin{align}
F&=(1.39 \times 10^{30} \mathrm{s}^{-1})(m_p/(10^{-24} \mathrm{g/cm^3}))\nonumber\\
&\quad\times(N(A_1,A_2)/V_\mathrm{gal})(M/M_\odot)(R/R_\odot)\, v_{250}^{-1}\nonumber\\ 
&=(3.4 \times 10^{-37}\mathrm{s}^{-1})N(A_1,A_2)(M/M_\odot)(R/R_\odot)\, v_{250}^{-1}\label{capture}
\end{align}
where $m_p$ is the mass of the proton, $v_{250}$ is the velocity of the strangelets divided by 250 km/s, $V_\mathrm{gal}\sim 7\times 10^{66}\, \mathrm{cm^3} $ is the volume of the galaxy and $N(A_1,A_2)=\int_{A_1}^{A_2} {\mathrm d}A\,Q(A)$.
The most interesting case is for proto-neutron stars:
during a time $\tau_\mathrm{melt}$  (the first months of their life) a solid crust has not yet formed (and therefore the strangelets do not get trapped into that low-density layer) and a strangelet can penetrate the star triggering the deconfinement of the neutron star into a QS \cite{Madsen:1989pg}. Only strangelets with $A>10^{12}$ are not trapped by the expanding supernova shell \cite{Madsen:1989pg}, but from our analysis we can conclude that almost all the strangelets satisfy this condition (see Fig.\ref{distributions}). The flux of strangelets during that period turns out to be: $F \,\tau_\mathrm{melt} \sim 5\times 10^{-4}\, v_{250}^{-1} \ll 1$ and therefore the probability of this mechanism is  negligible.
Eq.(\ref{capture}) suggests capture by main sequence stars has a rate  $\sim 1/{\rm yr}$, but strangelets would likely evaporate during the pre-supernova collapse due to the high temperatures reached, as shown in Fig.~\ref{timescaleevap}. Moreover, if $v_{250}\gg 1$  (as suggested by the high velocities of matter ejected in a NS-NS merger) they would spread on a volume much larger than $V_\mathrm{gal}$ further reducing the capture rate described by eq.~(\ref{capture}).

Finally, strangelets can be detected by a variety of experiments, see e.g. \cite{Adriani:2015epa,Adriani:2017bfx,Burdin:2014xma} and references therein. If we assume that the strangelets move at a velocity close to that of the galactic halo, i.e. $v_s\sim 250 \,\mathrm{km/s}$, the total flux of strangelets in the galaxy is:
\begin{equation}
dj_s/d\Omega = v_s N_\mathrm{tot}^\mathrm{gal} /(4 \pi V_\mathrm{gal} )\sim 8.4\times 10^{-30} \mathrm{cm^{-2} s^{-1} sr^{-1}}\, ,
\end{equation}
many orders of magnitude smaller than experimental limits (and again if $v_{250}\gg 1$ the strangelets can escape the galaxy and the flux can be even smaller). 

 Our results suggest that strangelets produced by a QS-QS merger (and most likely also those which are possibly ejected by a NS-QS merger) are unlikely to be detected directly by experiments \cite{Adriani:2015epa,Adriani:2017bfx} or indirectly through the effects they produce on stellar evolution \cite{Madsen:1989pg} i.e. converting all NSs  into QSs. We reached this conclusion because we have not assumed a distribution of the sizes of the strangelets of the form $\delta(A-\bar A)/A$ (i.e. all the strangelets having the same mass $\bar A$), as it is often done in the analysis, but we have considered an explicit model for fragmentation and evaporation. 

In our paper we have therefore shown that the existence of the QSs is not excluded by the present data on the abundance of not too massive strangelets produced in the merger of two QSs.
The possibility of having at least a fraction of dark matter composed of strangelets produced at the time of the baryogenesis and with $A$ significantly larger, in the range $\sim 10^{33}-10^{42}$, is also still viable \cite{,Witten:1984rs,Burdin:2014xma,Jacobs:2014yca,SinghSidhu:2019tbr}. The experimental search of strangelets of all possible masses is still a very active research field \cite{Bacholle:2020emk,POEMMA:2020ykm}.

\acknowledgements{Acknowledgements: A.B. acknowledge support by the European Research Council (ERC) under the European Union's Horizon 2020 research and innovation programme under grant agreement No. 759253 and by Deutsche Forschungsgemeinschaft (DFG, German Research Foundation) - Project-ID 279384907 - SFB 1245 and DFG - Project-ID 138713538 - SFB 881 (``The Milky Way System'', subproject A10).}

\begin{figure*}[t!]
\includegraphics[width=0.48\linewidth]{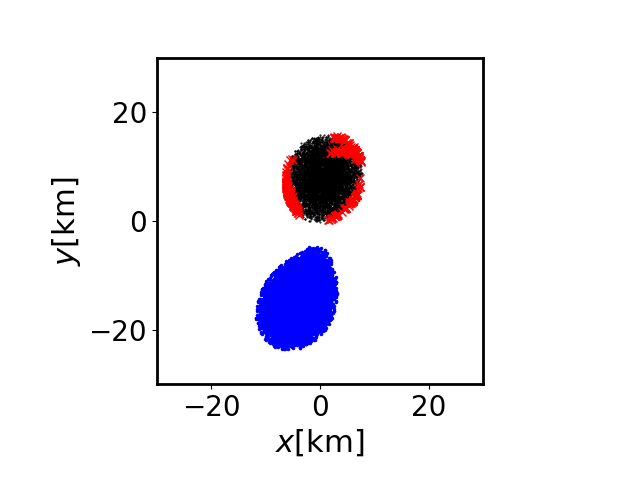}\includegraphics[width=0.48\linewidth]{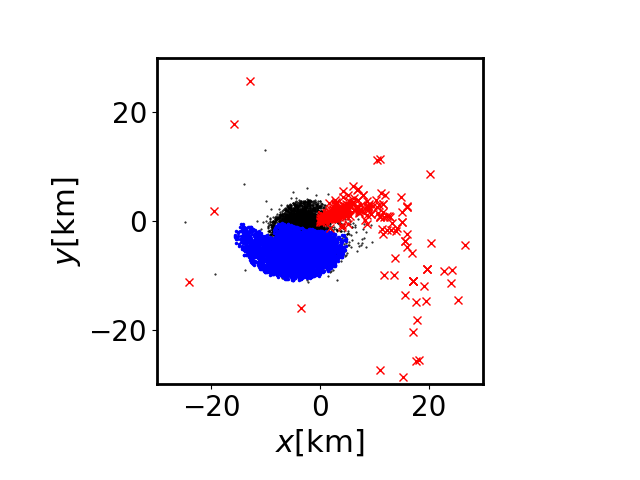}
\caption{Snapshots from a hydrodynamical simulation of a QS-NS merger just before (left) and after (right) merging. Blue dots indicate quark matter, while the NS is shown with black markers. (Note that we plot only a subset of SPH particles, which is why the stellar surfaces appear less smooth.) The red crosses mark fluid elements which eventually become unbound. Only hadronic matter is ejected.}\label{snap}
\end{figure*}

\section{Appendix A}
In order to assess the strangelets' production in a merger of a QS with a NS, we perform relativistic hydrodynamical simulations of these systems as in~\cite{Bauswein:2008gx,Bauswein:2009im}. Within the smoothed particle hydrodynamics (SPH) scheme it is relatively trivial to implement different equations of state (EoSs) for different fluid elements, i.e. SPH particles. We set up a binary with a NS of 1.35~$M_\odot$ and a QS with 1.35~$M_\odot$. The SPH particles of the NS are described by a hadronic EoS containing hyperons and deltas already used in the simulations presented in  \cite{DePietri:2019khb}, while we use a quark matter EoS for the QS based on the MIT bag model, again taken from \cite{DePietri:2019khb}. We run the simulations into the early merger phase and stop the calculations when a black hole forms a few milliseconds after merging for these particular binary masses and relatively soft EoSs  \footnote{It is possible that if a stiffer quark matter EoS is chosen, as e.g. the one used in \cite{Bombaci:2020vgw} the collapse to BH is postponed.}. We do not include any burning processes, as the ones described in~\cite{Herzog:2011sn,Drago:2015fpa} that could convert nucleonic matter to quark matter, because such a conversion is likely less relevant on short time scales, although it could become more important at later times. It is thus unlikely that neglecting these processes does change our conclusions about dynamical mass ejection.

Analyzing the mass ejection in this simulation we do not find any evidence for unbound quark matter, see Fig.~\ref{snap}. Instead only matter from the NS gets ejected because this material is less bound than the quark matter. For this particular set of EoSs, the hadronic one yields smaller stellar radii for stars with 1.35~$M_\odot$, which is why the QS initially shows stronger tidal deformations before merger, but still it does not eject quark matter. The hadronic matter is less bound and the NS gets in part disrupted during the merging process, whereas the quark matter component remains entirely gravitationally bound.

Note that the equal-mass binary is the most extreme case within such scenario of a ``mixed'' merger. Actually, one would expect that in most of the cases the QS is more massive than the NS \cite{DePietri:2019khb} implying that mass ejection from the QS is even more unlikely for asymmetric binaries. 

In another simulation we consider the merger of two stars with 1.2~$M_\odot$ which may also represent a rather extreme system favoring quark matter ejection. For these binary masses the merger does not lead to a direct gravitational collapse. We find a very small amount of unbound quark matter (less than $10^{-4} M_\odot$) compared to several $10^{-3}~M_\odot$ of ejecta from the NS. We remark that the small amount of quark matter ejecta corresponds to just a few SPH particles, and thus we cannot fully exclude that those are a numerical artifact. In this study we cannot provide a full survey of all possible binary masses and EoSs for mixed binaries. We cannot therefore fully exclude that other systems may lead to the ejection of quark matter, but it is reasonable to expect that even if in that case the thermodynamical and mechanical properties of the strangelets would be similar to those in QS-QS mergers, which we extensively discuss in the main body of the paper.

\section{Appendix B}
In this Appendix we sketch the derivation of Eq.(\ref{scaling}).
First, the probability
of interaction between the neutrino and the strangelet $p(r_s,T)$ reads:
\begin{eqnarray}
p&=&1\,\, \mathrm{for}\,\, r_s>\frac{3}{4}l(T)\nonumber\\
\,&=&\frac{4r_s}{3l(T)}\,\,\mathrm{for}\,\, r_s\leq\frac{3}{4}l(T)
    \end{eqnarray}
with the neutrino mean free path $l(T)=(G_F^2\mu^2 T^3)^{-1}$.

For small strangelets (as the ones discussed in this paper), $p$ scales therefore as $\sim r_s T^3$ and the LHS of Eq.(\ref{neutrini}) scales therefore as $A$. Eq.(\ref{neutrini}) expresses the thermal equilibrium between the cooling due to evaporation (RHS) and the re-heating due to neutrinos present in the ambience at a finite temperature $T_u$ (LHS). That equation can be combined with the explicit expression of the netto evaporation rate given by Eq.(\ref{evap}) to obtain:
\begin{equation}
k_1A(T_u^7-T_s^7)=k_2A^{2/3}B(T_u,T_s)(I+2T_s)    \label{equilib}
\end{equation}
where 
\begin{equation}
B(T_u,T_s)=k_3T_s^2e^{-I/T_s}-k_4\frac{T_u^4-T_s^4}{\sqrt{T_s}}
\end{equation}
and $k_1,k_2,k_3,k_4$ are constants.
Notice that the LHS of Eq.~(\ref{equilib}) describes the reheating due to neutrinos and it scales as $A$, while the RHS describes the evaporation and it scales as $A^{2/3}$. We can then obtain:
\begin{equation}
A^{1/3}=\frac{k_2}{k_1}\frac{I+2T_s}{T_u^7-T_s^7}f(T_s)   \, . \label{eqcompact}
\end{equation}
This equation describes the relation between the size of the strangelets and their temperature at equilibrium, if $T_u$ is given. It indicates that if the size of the strangelets is small ($A\rightarrow 0$) no evaporation is possible ($f(T_s)\rightarrow 0$ and therefore also $dA/dt\rightarrow 0$) because the re-heating is not efficient and the temperature of the strangelets drops.
Let us indicate with $T_0$
the value of $T_s$ for which $f(T_s)=0$ and thus also $dA/dt=0$.
For $T_s<T_0$ reabsorption dominates over evaporation.
Notice that $T_0$ depends on $T_u$ but does not depend on $A$.
Let us expand the LHS of Eq.(\ref{eqcompact}) around $T_0$ by introducing a small shift $\delta$: $T_s=T_0+\delta$.
It is then easy to check that $\delta \propto A^{1/3}$ and thus the netto evaporation rate reads:
\begin{equation}
\frac{dA}{dt} \propto A^{2/3}\delta =k A\,. 
\end{equation}
From this last equation the total evaporation time (needed to decrease the baryon number from $A$ to $1$) reads: 
\begin{equation}
\tau=\log(A)/k
\end{equation}
i.e. Eq.(\ref{scaling}) in the main text. We tested this equation against the numerical results and they are in excellent agreement.

\begin{figure}[b]
\includegraphics[width=\linewidth]{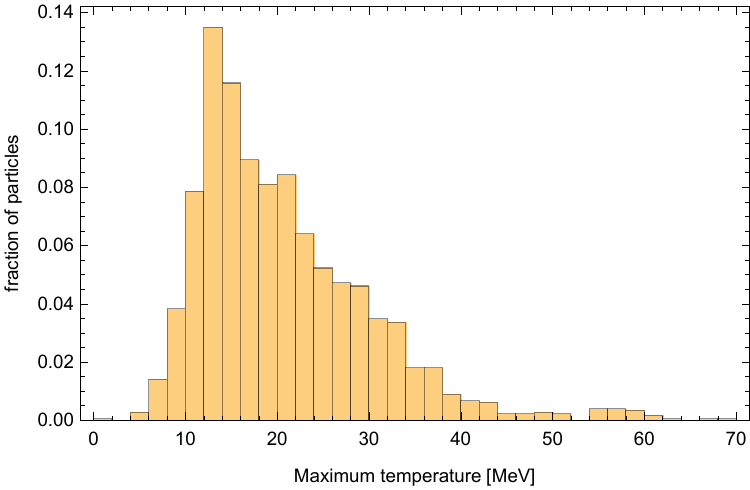}
\caption{Distribution of the maximum temperatures reached by each of the ejected particles.}\label{temperature}
\end{figure}

\section{Appendix C}
In this Appendix we outline the formalism which allows to describe how the total or partial evaporation of the ejected fragments of strange matter affects their size distribution.
During the first milliseconds after the beginning of the merger, fragments of quark matter not only scatter breaking into smaller pieces, but they also evaporate due to the high temperatures reached by the system and shown in Fig.~\ref{temperature}. We can then compare the temperatures reached by each of the ejected particles with the time needed for a total evaporation, displayed in Fig.~2 of the text. By using Eq.~(5): $\tau_A = C (T) \mathrm{Log} A$ which indicates the time needed for the total evaporation of a strangelet of baryon number $A$ and by numerically evaluating $C(T)$ we can obtain the estimate of the time needed for a partial evaporation:
\begin{equation}
\tau_A (f) = C (T) (\mathrm{Log} A - \mathrm{Log} (f A))= -C(T) \,\mathrm{Log} f\, , 
\end{equation}
where f is the fraction of the strangelet that has NOT evaporated. For instance f=0.01 indicates 99\% evaporation, leaving a strangelet having baryon number 0.01 A. Notice that $\tau_A (f)$ depends on $A$ only through the limit $f>1/A$, but it is otherwise $A$ independent as can be seen also from the results presented in Table~\ref{tabevap}. It is then possible to estimate the probability that each of the particles leaving the merger's region after the collision and having initial mass A evaporates at least up to a fraction $f$, meaning that the remaining not-evaporated mass of the particle has baryon number at most $f A$. One can consider each of the trajectories, following their thermal evolution, and check if that particle remained at a temperature $T$ for a time longer than $\tau_A (f)$, so to be able to evaporate up to the fraction $f$. In this way we obtain the probability that the ejected particles having initial baryon number $A$ evaporate at least up to the fraction $f$. We can consider this quantity $R_A(f)$ as a cumulative probability, related to the corresponding probability distribution function $E_A(f)$ by:
\begin{equation}
R_A(f)=\int_{1/A}^f \mathrm{d}f' E_A(f') + R_A(1/A) \, ,
\end{equation}
where $R_A(1/A)$ is the probability of total evaporation of a strangelet having baryon number $A$. We then extract the numerical value of the evaporation probability distribution from its cumulative, as:
\begin{equation}
E_A(f) = \partial R_A(f) / \partial f \, .
\end{equation}
We can now define $Q(A)$, the distribution of the size of the fragments after the evaporation in terms of $P(A)$, the distribution obtained from the scattering simulation:
\begin{align}
Q(A)&=\int_1^\infty \mathrm{d} A' P(A')\int_{1/A'}^1 \mathrm{d}f\, E_{A'}(f) \delta(A-f A') \nonumber \\
     &= \int_1^\infty \mathrm{d} A' P(A')\, E_{A'}(f=A/A')/A' \, .
\end{align}
This definition satisfies the number of fragments sum-rule:
\begin{equation}
\int_1^\infty \mathrm{d} A\, Q(A) = \int_1^\infty \mathrm{d} A \,P(A) [1-R_A(1/A)]
\end{equation}
indicating that the number of fragments after evaporation is equal to the number of fragments before evaporation minus the number of fragments that underwent total evaporation. It is important to remark that the total number of fragments, i.e. of strangelets having $A>1$, can only be reduced by evaporation and it is therefore limited by the number of strangelets produced by fragmentation.

\begin{table}[H]
\begin{center}
\begin{tabular}{||c | c | c | c||}
\hline
& 100 $\%$ & 99.9 $\%$ & 99 $\%$\\
&  evaporated & evaporated & evaporated\\
\hline
A=$\rule{0pt}{4ex}  10^{10}$ & 0.68 &  0.78 & 0.81\\
A=$10^{35}$ & 0.57 &  0.78 & 0.81\\
\hline
\end{tabular}
\end{center}
\caption{Probability of partial or total evaporation of a strangelet having a baryon number either of $10^{10}$ or of $10^{35}$, assuming a binding energy $I=70$ MeV. The temperatures reached by the ejected particles are obtained from the simulation of a 
QS-QS merger of Ref.\cite{Bauswein:2009im}, using the equation of state MIT60 with masses $1.2$ and $1.35 M_{\odot}$.}\label{tabevap}
\end{table}

\bibliography{references1}

\end{document}